\documentclass[twocolumn,preprintnumbers,amsmath,amssymb]{revtex4}
\usepackage{graphicx}
\usepackage{dcolumn}
\usepackage{bm}

\newcommand{\comment}[1]{}
\newcommand{\bdp}{B^{\prime \prime}_z}
\newcommand{\bp}{B^\prime}
\newcommand{\bs}{B_s}
\newcommand{\brot}{B_{\mbox{\scriptsize \textit{rot}}}}
\newcommand{\bdpeff}{B^{\prime \prime}_{\mbox{\scriptsize \textit{eff}}}}

\begin{document}

\title{Bose-Einstein condensation in a circular waveguide}

\author{S.\ Gupta, K.W.\ Murch, K.L.\ Moore, T.P.\ Purdy, and D.M.\ Stamper-Kurn}
\affiliation{Department of Physics, University of California,
Berkeley CA 94720}

\date{\today }

\begin{abstract}
We have produced Bose-Einstein condensates in a ring-shaped
magnetic waveguide. The few-millimeter diameter non-zero bias ring
is formed from a time-averaged quadrupole ring.  Condensates which
propagate around the ring make several revolutions within the time
it takes for them to expand to fill the ring.  The ring shape is
ideally suited for studies of vorticity in a multiply-connected
geometry and is promising as a rotation sensor.
\end{abstract}


\maketitle


Scalar superfluids  are characterized by a complex order parameter
$\Psi(\bf{r})$ which is uniquely defined throughout the fluid.
This implies the irrotational motion of the fluid in the space
where $\Psi(\bf{r}) \neq 0$, leading to the Meissner effect in
charged superfluids and to the Hess-Fairbank effect in neutral
ones.  Given this constraint, rotational motion of superfluids (or
magnetic flux density in type-II superconductors) is accommodated
by lines of quantized vorticity which disrupt the simple
connectivity of the fluid. Multiple connectivity can also be
imposed by the proper design of containers for the fluids. Such
geometries are enlisted to translate phase variations of
$\Psi(\bf{r})$ into sensors of external fields. For example,  a
SQUID magnetometer makes use of a superconducting ring interrupted
by Josephson junctions  to allow continuous sensitivity to
magnetic fields. A similar geometry was used in a superfluid
$^3$He gyroscope \cite{schw97}.

Dilute gas superfluids enable novel forms of matter-wave
interferometry. Precise sensors of rotation, acceleration, and
other sources of quantal phases \cite{petr99,lyan00} using trapped
or guided atoms have been envisioned.  In particular, the
sensitivity of atom-interferometric gyroscopes is proportional to
the area enclosed by the closed loop around which atoms are guided
\cite{gust00sagnac}. Such considerations motivate the development
of closed-loop atom waveguides which enclose a sizeable area.

A number of multiply-connected trapping geometries for cold atoms
have been discussed.  Optical traps using high-order
Gauss-Laguerre beams were proposed \cite{wrig00,ande03}, and
hollow light beams were used to trap non-degenerate atoms in an
array of small-radius rings \cite{verk03}. Large-scale magnetic
``storage rings'' were developed for cold neutrons \cite{kugl85}
and discussed for atomic hydrogen \cite{thom89}. More recently,
closed-loop magnetic waveguides were demonstrated for laser cooled
atoms \cite{saue01,wu04guide}. Unfortunately, these  guides are
characterized by large variations in the potential energy along
the waveguide and by high transmission losses at points where the
magnetic field vanishes.

In this Letter, we report the creation of a smooth, stable
circular waveguide for ultracold atoms.  A simple arrangement of
coaxial electromagnetic coils was used to produce a static
ring-shaped magnetic trap, which we call the quadrupole ring
(Qring), in which strong transverse confinement is provided by a
two-dimensional quadrupole field.  Atoms trapped in the Qring
experience large Majorana losses, but we can eliminate such losses
with a time-orbiting ring trap (TORT) \cite{arno04tort}. In this
manner, stable circular waveguides with diameters ranging from 1.2
to 3$\,$mm were produced. Finally, we report on the production of
Bose-Einstein condensates (BECs) in a portion of the circular
waveguide, and on the guiding of an ultracold atomic beam for
several revolutions around the guide. This ring-shaped  trap
presents opportunities for studies of BECs which are homogeneous
in one dimension and therefore of the unterminated propagation of
sound waves \cite{andr97prop} and solitons
\cite{sala99toroid,bran01,mart01}, of persistent currents
\cite{bloc73,java98pers,bena99,nuge03}, of quantum gases in low
dimensions, and of matter-wave interferometry.

To explain the origin of the quadrupole ring trap, we consider a
cylindrically-symmetric static magnetic field $\vec{B}_c$ in a
source-free region.  Expanding $\vec{B}_c$ to low order about a
point (taken as the origin) on the axis where the field magnitude
has a local quadratic minimum, we have
\begin{equation}
\vec{B}_{c} = B_0 \hat{z} + \frac{\bdp}{2} \left[ \left( z^2 -
\frac{x^2 + y^2}{2}\right) \hat{z} - z \left( x \hat{x} + y
\hat{y} \right) \right] \label{eq:bcurv}
\end{equation}
where $B_0\!>\!0$ is the field magnitude at the origin, $\bdp$ the
axial field curvature, and with cartesian coordinates
($x$,$y$,$z$) chosen so that $z$ is the axial coordinate.  The
magnetic field magnitude falls to zero in the $\hat{x}$-$\hat{y}$
plane along a circle of radius $\rho_0 = 2 \sqrt{B_0 / \bdp}$
centered at the origin.  This is the Qring, a  ring-shaped
magnetic trap for weak-field seeking atoms. Near the field zeros,
the magnetic field has the form of a transverse (radial and axial
directions) two-dimensional quadrupole field with gradient $\bp
=\sqrt{B_0 \bdp}$.  Such traps can also be obtained using
different electromagnet configurations \cite{arno04tort}.

\begin{figure}
\includegraphics[angle = 0, height = 1.75 in] {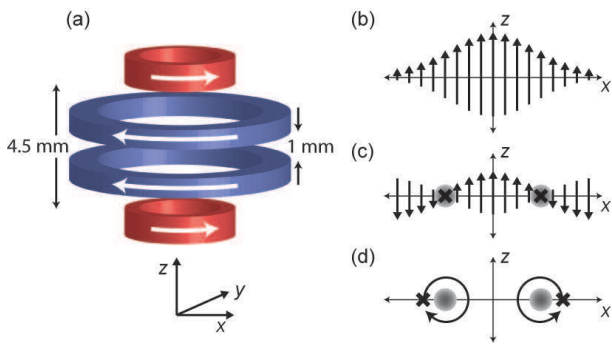}
\caption{\label{scheme} Forming a circular magnetic waveguide. (a)
Four coaxial circular electromagnets (see \cite{moor05} for
details) are used to generate both the static (currents as shown)
and rotating fields needed for the waveguide. Axes are indicated;
gravity points along $- \hat{z}$. (b) As shown schematically, the
field (arrows) from just the two outer coils (curvature coils,
colored red) points axially in the midplane between the coils,
with largest fields at the axis. (c) Adding a uniform opposing
bias field (using anti-bias coils, colored blue) produces a ring
of field zeros (X) in the $\hat{x}$-$\hat{y}$ plane about which
weak-field seeking atoms (shaded region) are trapped. (d) Rapidly
rotating the field zeros around the trapped atoms produces the
TORT.}\label{fig:scheme}
\end{figure}

In our apparatus, the Qring is  formed using a subset of the coils
(the curvature and anti-bias coils, see Fig.\ \ref{fig:scheme})
used in our recently demonstrated millimeter-scale Ioffe-Pritchard
magnetic trap \cite{moor05}. Our work is aided in particular by
the large axial curvatures produced in this trap and by the
vertical orientation of the trap axis. These features are relevant
for the operation of a Qring in the presence of gravity, for two
reasons. First, trapping atoms in the Qring requires transverse
confinement sufficient to overcome the force of gravity; this
places a lower bound  on the radius of the Qring of $\rho_0\! > \!
\rho_{min} \!\simeq \! 2 m g / |\mu| \bdp$ with $m$ the atomic
mass, $g$ the acceleration due to gravity, and $\mu$ the atomic
magnetic moment. Indeed, if $\rho_{min}$ exceeds the range over
which Eq.\ \ref{eq:bcurv} is valid, typically the distance to the
field-producing coils, the formation of a Qring may be precluded
entirely.  With $\bdp = 5300$ G/cm$^2$ in our experiments,
$\rho_{min} = 115 \, \mu$m is much smaller than the millimeter
dimensions of the electromagnets used for the trap. Second, the
vertical orientation of the Qring axis allows cold atoms to move
slowly along the nearly horizontal waveguide rather than being
confined in a deep gravitational well.

Atoms can be localized to a particular portion of the Qring by
application of a uniform sideways (in the $\hat{x}$-$\hat{y}$
plane) magnetic field; e.g.\ a weak bias field $\bs \hat{x}$ tilts
the Qring by $\Delta z/\rho_0 = (\bs / \bp)/ \rho_0$ about the
$\hat{y}$ axis. This adjustment also adds an azimuthal field of
magnitude $\bs |\sin \phi|$, splitting the Qring into two trap
minima at opposite sides of the ring, with $\phi$ being the
azimuthal angle conventionally defined.

\begin{figure}
\includegraphics[angle = 0, height = 2.1 in] {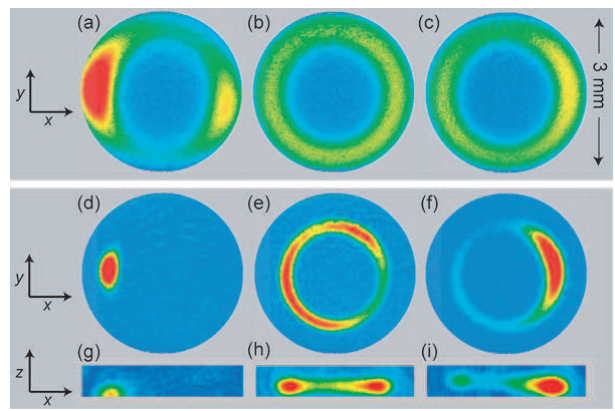}
\caption{\label{fig:qringimage} Atoms in a ring-shaped magnetic
trap.  Shown are top (a-f) and side (g-i) absorption images of
ultracold $^{87}$Rb clouds in either a Qring (a-c) or TORT (d-i)
with applied side field $\bs$= 9.2 (left), 0 (middle) and
-2.5$\,$G (right column) respectively in the $\hat{x}$ direction.
Images were taken 2 ms after turning off the traps. The applied
field tilts the Qring or TORT and favors atomic population in one
side or another of the trap. For $\bs \sim 0$, the trap lies
nearly in the horizontal plane and its azimuthal potential
variation is minimized. For the Qring, $B_0 = 22$ G; for the TORT,
$B_0 = 20$ G and $\brot = 17$ G; and $\bdp = 5300$ G/cm$^2$ for
both.  The temperature of trapped atoms is 90 $\mu$K in the Qring,
and 10 $\mu$K in the TORT.  On-resonance absorption ranges from 0
(blue) to $>80\%$ (red).}
\end{figure}

We loaded cold atoms into the Qring using a procedure similar to
previous work \cite{moor05}. Briefly, about $2 \times 10^9$
$^{87}$Rb  atoms in the $|F=1, m_F = -1\rangle$ hyperfine ground
state were loaded into one of two adjacent spherical quadrupole
magnetic traps. Using these traps, atoms were transported 3 inches
from the loading region to the Qring trap region. During this
transport, RF evaporative cooling was applied, yielding $2.5
\times 10^7$ atoms at a temperature of $60\,\mu$K in a spherical
quadrupole trap with an axial field gradient of 200 G/cm. Within 1
s, we then converted the spherical quadrupole to a tilted-Qring
trap produced with $\bdp = 5300$ G/cm$^2$, $B_0 = 22$ G, and a
side field of magnitude $\bs = 9.2\,$G. This process left $2
\times 10^7$ atoms trapped in the Qring (Fig.\
\ref{fig:qringimage}).


The trapping lifetime of atoms in the Qring is limited by Majorana
losses. In a balanced Qring, trapped atoms passing close to the
line of zero field, which extends all around the ring, may flip
their spins and be expelled from the trap. Extending the treatment
by Petrich \emph{et al.}\ \cite{petr95} to this scenario, we
estimate a Majorana loss rate of $\frac{\hbar^{1/2}}{\pi m^{3/4}}
\frac{(\mu \bp)^{3/2}}{(k_B T)^{5/4}} = 6 \, \mbox{s}^{-1}$ for
our trap at a temperature of 60 $\mu$K. In a tilted Qring, the
zero-field region is reduced to just two points at opposite sides
of the ring. Majorana losses in a tilted Qring are thus similar to
those in spherical quadrupole traps and much smaller than in a
balanced Qring.  We confirmed this qualitative behaviour by
measuring the lifetime of trapped atoms in balanced and tilted
Qring traps. In the balanced Qring, the measured 0.3 s$^{-1}$
Majorana loss rate was thrice that in a tilted Qring, while
falling far short of the predicted 6 s$^{-1}$ loss rate,
presumably due to residual azimuthal fields.

The high loss rates in the Qring can be mended in a manner similar
to the time-orbiting potential (TOP) trap by which Majorana losses
in a spherical quadrupole field were overcome \cite{petr95}. As
proposed by Arnold \cite{arno04tort}, a time-orbiting ring trap
(TORT) with non-zero bias field  can be formed by displacing the
ring of field zeros away from and then rapidly rotating it around
the trapped atoms (Fig.\ \ref{fig:scheme}(c)). From Eq.\
\ref{eq:bcurv}, the Qring can be displaced radially  by
application of an axial bias field, and displaced along $\hat{z}$
by a cylindrically-symmetric spherical quadrupole field.  The TORT
provides transverse quadratic confinement with an effective field
curvature of $\bdpeff = {\bp}^2 / 2 \brot$, where $\brot$ is the
magnitude of the rotating field seen at the trap minimum. Just as
the TOP trap depth is limited by the ``circle of death,'' the TORT
trap depth is limited by a ``torus of death,'' the locus of points
at which the magnetic field is zero at some time
\cite{evapfootnote}.   This scheme may be applied equally to a
tilted Qring, yielding a tilted TORT and providing a stable trap
in which atoms are confined to a portion of the ring.  The
sideways magnetic field (e.g.\ along $\hat{x}$) causes the
magnetic potential minimum to vary azimuthally in the tilted TORT
as $|\mu| \left(\brot^2 + \bs^2 \sin^2\phi\right)^{1/2}$.  The
variation in the gravitational potential is the same as that in a
Qring.

The time-varying fields needed to convert our Qring (or tilted
Qring) traps to TORT (or tilted TORT) traps were obtained by
suitably modulating the currents in the four coils used to
generate the Qring potential. A modulation frequency of $5$ kHz
was chosen to be much larger than the transverse motional
frequencies ($<100\,$Hz) and also much smaller than the Larmor
frequency ($> 3\,$MHz) at the location of the trap minimum.  To
first switch on the TORT, a rotating field magnitude of $\brot =
18$ G was used \cite{brotfootnote}.

\begin{figure}
\includegraphics[angle = 0, height = 1.75 in] {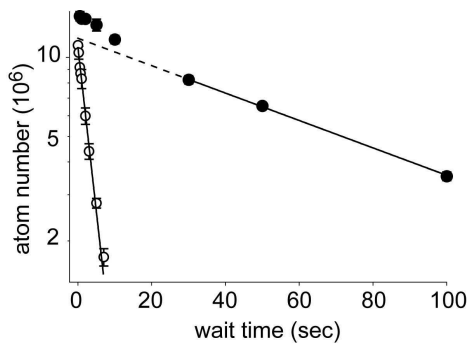}
\caption{\label{fig:lifetime} Elimination of Majorana losses in
the TORT. The measured number of trapped atoms in a Qring (open
circles) or TORT (filled circles) trap is shown vs.\ residence
time in the trap. Exponential fits indicate a 3~s
Majorana-loss-limited lifetime in the Qring.  In the TORT,
following an initial (30 s) loss of atoms due to evaporation, a
vacuum limited lifetime of 90 s was observed.  Settings for $B_0$,
$\bdp$ and $\brot$ are as in Fig.\ \ref{fig:qringimage}, $\bs \sim
0$, and the initial temperature is 60 $\mu$K.}
\end{figure}

As shown in Fig.\ \ref{fig:lifetime}, the trap lifetime was
dramatically increased by application of the TORT trap. In the
first few seconds after switching on the TORT, we observed a fast
loss of atoms and a simultaneous drop in their temperature. We
ascribe this loss and cooling to the evaporation of atoms from the
trapped cloud through the ``torus of death.''   As the temperature
dropped, the evaporation rate diminished and lifetime of trapped
atoms became vacuum limited at 90 s, a value observed both for
balanced and for tilted TORT traps.

Given their longevity, it is possible to evaporatively cool
TORT-trapped atoms to the point of quantum degeneracy. Using a
tilted TORT with $\bs \sim 9\,$G, evaporation was performed in two
stages.  First, ``torus of death'' evaporation was applied by
ramping down the rotating field strength $\brot$ over 40$\,$s to
$4.8\,$G. The oscillation frequencies in this trap were measured
as $\omega_\perp = 2 \pi \times (87,74.5)\,$Hz in the transverse
and $\omega_\phi = 2 \pi \times 35\,$Hz in the azimuthal
directions, in agreement with predictions
\cite{transversefootnote}.  In the second stage, RF evaporation
was applied for 20$\,$s, yielding clouds of up to $6 \times 10^5$
atoms at the Bose-Einstein condensation temperature, and pure BECs
of up to $3 \times 10^5$ atoms.


\begin{figure}[t]
\includegraphics[angle = 0, height = 3in] {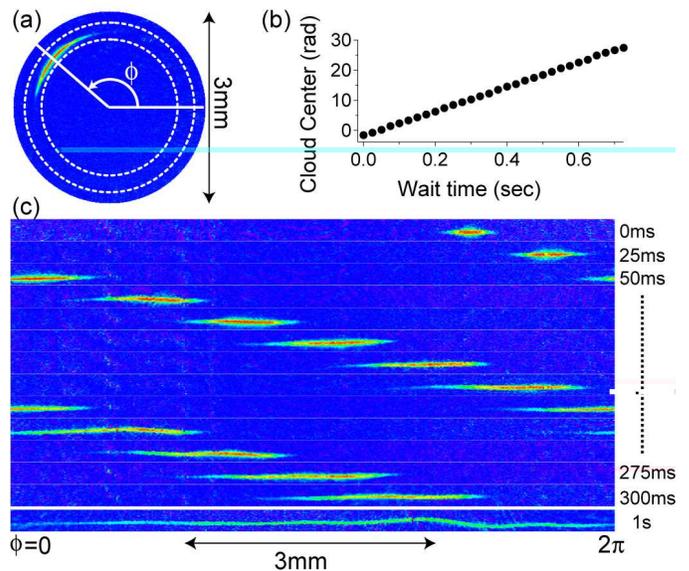}
\caption{\label{fig:around} Circular motion of a quantum
degenerate atomic beam in a waveguide.  A Bose-Einstein condensate
was launched into a balanced TORT and allowed to propagate. (a)
Top view in-trap absorption image during the propagation. The mean
azimuthal position of the BEC measured from such images is shown
in (b).  Annular portions (indicated by dashed circles) of
top-view images taken at different guiding times are shown in (c)
displayed in polar coordinates (radius vs.\ azimuthal angle).  The
beam advances at an angular velocity of 40.5 rad/s while expanding
due to an rms azimuthal velocity spread of 1.4 mm/s. After 1 s,
the beam expands so as to fill the entire guide.}
\end{figure}

Finally, to assess the suitability of the TORT as an atomic
waveguide for interferometry, we launched our trapped BECs into
closed-loop circular motion along the guide. This was accomplished
by reorienting the sideways bias field $\bs$, inducing the trapped
BEC to accelerate toward the newly-positioned tilted TORT trap
minimum (advanced by an azimuthal angle of about $\pi/4$), while
simultaneously reducing the magnitude of $\bs$ to $\bs \sim 0$ and
increasing $\brot$ to 12.6 G to produce a well-balanced TORT trap.
The TORT was then maintained at this setting, with radius $\rho_0
= 1.25$ mm ($B_0 = 20$ G, $\bdp = 5300$ G/cm$^2$), and transverse
trap frequencies of $\omega_\perp \simeq 2 \pi \times 50$ Hz as
measured at the launch-point of the atoms. The atoms were allowed
to propagate freely around the guide for various guiding times
before being observed by absorption imaging. As shown in Fig.\
\ref{fig:around}, the ultracold atomic beam propagated  around the
circular waveguide at an angular (linear) velocity of 40.5 rad/s
(50.6 mm/s).  As measured from the the azimuthal extent of the
atoms for different guiding times, this pulsed atom beam was
characterized by an azimuthal rms velocity spread of 1.4 mm/s,
equivalent to a longitudinal temperature of 22 nK. After about 1 s
of guiding, this velocity variation caused the atomic cloud to
spread throughout the waveguide, by which point the atoms had
travelled $L\! =\! 51$ mm along the waveguide, encompassing an
area of $A\! =\! L \rho_0 / 2\! =\! 32$ mm$^2$.

Many requisite elements for interferometric rotation sensing are
still lacking in our system, including a means of in-guide
coherent atomic beam splitting \cite{wang05inter,wu05debroglie},
bidirectional propagation, proper radial waveguiding
\cite{ande02multi}, full characterization of longitudinal
coherence in the beam, an assessment of the influence of the
time-orbiting field on sensor precision,  and atom-interferometric
stability. Nevertheless, it is valuable to consider the possible
sensitivity of our system if these elements are attained. As
limited by atomic shot-noise, rotation measurements with an
uncertainty of $\Delta \Omega = (\hbar / 4 m A) N_0^{-1/2} \sim 1
\times 10^{-8}$ rad/s could be made from a single (1 s long)
measurement, where $N_0 = 3 \times 10^5$ is the number of atoms
used.  While this figure is nearly 20 times that of existing
atom-based gyroscopes \cite{gust00sagnac}, improvements such as
launching the atoms at higher velocities, increasing the TORT
radius, and increasing the atom number may ultimately yield a
useful, compact sensing device.

Other applications of the TORT may include studies of propagation
\cite{bong01guide,lean02guide,fort03guide} and non-linear dynamics
\cite{ott03guide} in atomic waveguides. In a TORT potential which
is modified either by application of magnetic fields or by tilting
with respect to gravity, BEC's can be studied both undergoing
pendular motion (terminated guide) when launched at small
velocities, and undergoing circular motion (unterminated guide) at
larger velocities.

Another appealing possibility is the study of BECs which fill the
ring-shaped trap, rather than forming in just a portion of the
ring. Such a system would allow for studies of quantized and
persistent circulation \cite{bloc73,java98pers,bena99,nuge03},
unterminated motion of solitons \cite{sala99toroid,bran01,mart01},
and other aspects of non-linear dynamics \cite{gara00hole}.  For
this purpose, the azimuthal variation in the TORT potential must
be reduced below the typical $\sim 100$ nK scale of the BEC
chemical potential.  From measurements of the kinetic energy of
BEC's undergoing circular motion in our trap (data of Fig.\
\ref{fig:around}), we estimate the TORT potential varied by as
much as 5 $\mu$K.  We believe this figure can be reduced greatly
by using traps of smaller radius, by designing better
electromagnets both for the static and the rotating fields used
for the TORT, and also by controlling the orientation of the
electromagnets with respect to gravity.

This work was sponsored by DARPA (Contract No.\ F30602-01-2-0524),
the NSF (Grant No.\ 0130414), the David and Lucile Packard
Foundation, and the University of California.  KLM acknowledges
support from the NSF, and SG from the Miller Institute.

\bibliographystyle{prsty}

\begin{thebibliography}{10}
\bibitem{evapfootnote}
Due to this topology, RF evaporation should be equally effective
whether one
  approaches the Larmor frequency at the trap minimum from above or below.
\bibitem{transversefootnote}
The transverse asymmetry arises due the sideways field $\bs$. In
the
  tilted-TORT, considering both magnetic and gravitational potentials, the
  azimuthal trap frequency is $\omega_\phi=\left(
  \left[mg\bs/\bp+\mu\bs^2/\brot\right]/m \rho_0^2\right)^{1/2}$.
\end{thebibliography}

\end{document}